\documentclass[aps,prl,reprint,amsmath,amssymb,superscriptaddress,notitlepage]{revtex4-1}
\usepackage[utf8]{inputenc}
\usepackage{bm}
\usepackage{enumerate}
\usepackage{graphicx}
\usepackage{xcolor}
\usepackage{hyperref}
\usepackage{dsfont}

\newcommand*{\twomatrix}[4]{\left[\begin{array}{cc}
    #1 & #2\\
    #3 & #4
  \end{array}\right]}

\begin{abstract}
Recent experimental searches for signatures of Majorana-like excitations in proximitized semiconducting nanowires involve conductance spectroscopy, where the evidence sought after is a robust zero-bias peak (in longer wires) and its characteristic field-dependent splitting (in shorter wires).
Although experimental results partially confirm the theoretical predictions, commonly observed discrepancies still include (i) a zero-bias peak that is significantly lower than the predicted value of $2e^2/h$ and (ii) the absence of the expected ``Majorana oscillations'' of the lowest-energy modes at higher magnetic fields.
Here, we investigate how the inevitable presence of a normal drain lead connected to the hybrid wire can affect the conductance spectrum of the hybrid wire.
We present numerical results using a one-band model for the proximitized nanowire, where the superconductor is considered to be in the diffusive regime, described by semi-classical Green functions.
We show how the presence of the normal drain could (at least partially) account for the observed discrepancies, and we complement this with analytic results providing more insights in the underlying physics.
\end{abstract}

\begin{document}

\title{Conductance spectroscopy on Majorana wires and the inverse proximity effect}

\author{Jeroen Danon}
\affiliation{Department of Physics, NTNU, Norwegian University of Science and Technology, 7491 Trondheim, Norway}

\author{Esben B. Hansen}
\author{Karsten Flensberg}
\affiliation{Center for Quantum Devices and Station Q Copenhagen, Niels Bohr Institute, University of Copenhagen, Copenhagen 2100, Denmark}

\date{\today}

\maketitle

Edges of bulk-gapped topological superconductors host localized zero-energy excitations that are commonly referred to as ``Majorana modes''~\cite{Volovik1999,Kitaev2001}.
These modes have non-Abelian anyonic braiding properties and could thus be used to implement fault-tolerant topological quantum computation~\cite{PhysRevLett.86.268,RevModPhys.80.1083}.
This notion sparked an intense search for systems that can host such excitations, and one attractive proposal is to use proximitized quasi-one-dimensional semiconducting nanowires~\cite{PhysRevLett.105.077001,PhysRevLett.105.177002,Sau2010}.
In the presence of strong enough spin-orbit interaction and induced $s$-wave superconducting pairing, the application of a magnetic field in a direction perpendicular to the effective spin-orbit field can induce a topological phase transition in such a wire, after which it behaves effectively as a gapped topological superconductor with two low-energy Majorana modes localized at the wire's ends.
%A network of such wires could in principle host a large number of Majorana modes, which could be braided by physically moving the excitations around using electric fields~\cite{alicea:natphys}, or by adiabatically manipulating inter-Majorana coupling strengths~\cite{PhysRevB.84.094505} or other wire parameters~\cite{Halperin2012}.

What makes this idea particularly attractive is the fact that all required ingredients rely in principle on well-established experimental techniques.
The proposal was thus rapidly followed by experiments, which used tunneling spectroscopy into one end of such a hybrid wire to detect the emergence of a zero-energy Majorana mode at high enough magnetic field~\cite{Mourik25052012,Deng2012a,das:natphys,PhysRevB.87.241401}.
Although a field-dependent zero-bias anomaly in the conductance was indeed a commonly observed phenomenon in these early experiments, several other observations were less compatible with an interpretation in terms of an emerging topological phase.
These ``inconsistencies'' included the zero-bias peak in the differential conductance being much smaller than the predicted value of $2e^2/h$ and the absence of a clear gap closing at the phase transition.

The experiments were thus immediately followed by a wave of theoretical work aimed at understanding the discrepancies.
Explanations that were consistent with having Majorana-like modes at the ends of the wire~\cite{Prada2012,Rainis2013,Mishmash2016} as well as alternative ``trivial'' interpretations of the observed zero-bias features~\cite{Liu2012b,Kells2012,Lee2012,Bagrets2012,Pikulin2012,Lee2013} were put forward.
In parallel, other ``smoking-gun'' features in the conductance spectrum were identified that could evidence a transition to a topological phase, a good candidate being the splitting of the zero-bias peak and subsequent characteristic oscillations of the low-energy modes as a function of magnetic field, due to finite-size effects~\cite{DasSarma2012,Rainis2013}.

In the years that followed, the quality of the experiments has steadily improved, mainly driven by advances in growth and fabrication techniques~\cite{Krogstrup2015}.
Today, state-of-the-art experiments~\cite{Albrecht2016,Deng2016,Zhang2016} show quite compelling evidence for the existence of Majorana modes in these hybrid nanowires, but a few annoying discrepancies persist:
(i) It is still very hard to measure a zero-bias peak that approaches $2e^2/h$ over a significant range of magnetic fields.
(ii) In most experiments the expected ``Majorana oscillations'' as a function of magnetic field are absent.
Recent experiments in the Coulomb-blockaded regime showed some oscillations~\cite{Albrecht2016}, but several of their characteristics do not fit current theory very well.
(iii) The zero-bias peak is usually much broader than expected, often filling most of the (quite soft) topological gap.

Several recent theoretical works addressed these points and investigated many effects in detail, including the occupation of multiple subbands in the wire, finite temperature, the existence of low-energy Andreev bound states in the wire~\cite{Chiu2017,Liu2017}, electrostatic interactions between the electrons in the wire and the substrate~\cite{Dominguez2016}, and having a finite subgap density of states in the proximitizing superconductor~\cite{Liu2017a,Stenger2017}.
A general trend is that the more ingredients are added to the model the better the theory can be made to resemble the experimental observations.

Here, we focus on one particular ingredient present in most experiments, which has been addressed only indirectly so far.
Inspired by the difference in behavior of the wires in the Coulomb-blockaded regime (where Majorana-like oscillations were observed) and in the transport regime (where oscillations are mostly absent), we propose that the presence of a second normal metal contact, usually connected as a drain lead to the superconductor, is a part of the setup that should be taken seriously.
Depending on the strength of the coupling between this drain and the wire (weak in the blockaded regime, stronger in a transport setup), the drain can induce a finite subgap normal density of states in the hybrid wire, a phenomenon known as the inverse proximity effect.
The bound states in the wire, including the low-energy Majorana modes, can thus acquire a finite life time which can be expected to affect the appearance of the measured conductance spectrum.
A crude way to account for this ``leakage'' is to simply add an imaginary part to the electronic energies in the superconductor, resulting in a broadening of all levels~\cite{Liu2017a,Stenger2017}.
Although this does produce a finite subgap density of states in the system, it does not provide a straightforward way to investigate any details related to the device geometry or the nature of the coupling between the drain and the wire.

\begin{figure}[t]
	\begin{center}
		\includegraphics[scale=1.2]{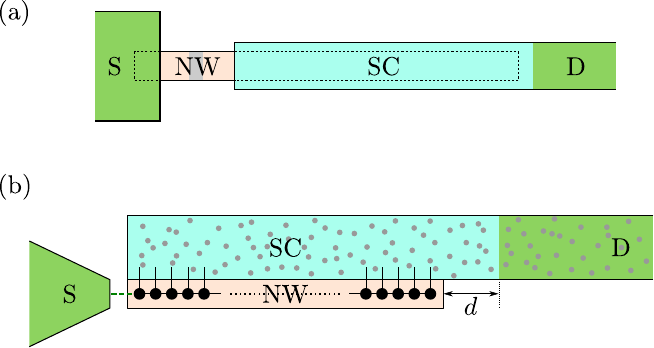}
		\caption{(a) Schematic of a commonly used setup for conductance spectroscopy experiments.
		The nanowire (marked `NW') is contacted on one side by a tunnel probe (marked `S') and is (partly) covered by an epitaxially grown $s$-wave superconductor layer (marked `SC'), which proximitizes part of the wire.
		This superconducting layer is connected to a normal metal drain lead (marked `D') through which the conductance is measured.
		(b) Sketch of the tight-binding model used for numerical calculations.
		The electron dynamics in the wire are discretized on a finite chain and the presence of the superconductor is included by adding a self-energy term in the electronic Green function.
		We treat the superconductor as being part of a diffusive SN-junction, where the junction interface is located a distance $d$ away from the nanowire.}
		\label{fig:modelsb}
	\end{center}
\end{figure}

Below, we present a detailed theoretical investigation of the effects of such a drain contact and we show how the results can indeed differ qualitatively depending on the geometry of the device and on the coherence properties of the superconductor.
The setup we will mainly have in mind is shown in Fig.~\ref{fig:modelsb}(a):
A semiconducting nanowire is proximitized by an epitaxially grown thin layer of superconductor, shown in blue.
A tunnel barrier (gray) in an uncovered part of the wire at the left end connects the proximitized region to a tunnel probe (the source contact `S').
A second normal lead (marked `D') is directly deposited onto the superconducting layer and serves as drain for transport measurements.

We model electronic transport in the superconductor (and in the drain lead) as being diffusive.
Although the actual mean free path in the superconductor (the distance between impurity scattering events) is probably not much shorter than all relevant device dimensions in most experiments, the surface of the superconductor on the outside (which usually forms an interface with an oxide layer) is known to be rough and can be expected to randomize the electrons' momentum each time they scatter off this surface.
We thus assume that we can use the thickness of the superconductor (typically 5--20~nm) as effective mean free path, which can justify employing a diffusion approximation.
Strictly speaking, the dynamics in such a thin layer are not necessarily exactly equivalent to the dynamics in a bulk diffusive medium, but we believe that it presents a reasonable approximation.

The rest of this paper is organized as follows:
We will first present a numerical study of the conductance spectrum of the system, where we treat the superconductor and the drain lead together as one diffusive SN-junction.
We will show how an efficient SN-coupling can be responsible for a suppression of the Majorana oscillations in the conductance spectrum.
The result is either a persistent zero-bias peak approaching $2e^2/h$ everywhere or a gradual suppression and smearing out of all features in the spectrum, depending on the ratio of the coherence length in the superconductor to the length of the proximitized region in the wire.
We also investigate the apparent hardness of the gap on both sides of the phase transition and find that a strong influence of the normal drain tends to soften the gap in the topological regime.
We then present a toy model where we only focus on the dynamics of the two low-lying (Majorana) modes.
We assume the modes to be coupled to each other and also include an effective coupling of both modes to the states in the source and drain leads.
From this simple model we derive an analytic expression for the differential conductance from source to drain.
We show how this result allows to qualitatively reproduce the main findings from our numerical calculations and we explain how it provides more insight in the underlying physics.

\section{Numerical tight-binding simulations}

We perform numerical tight-binding simulations of the conductance spectrum of a proximitized semiconducting nanowire, following the method we outlined in~\cite{Hansen2016}.
The main difference from earlier studies is that the superconductor is now treated as being part of a diffusive SN-junction and we use approximate expressions for the semi-classical regular and anomalous Green functions in this junction~\cite{Belzig1996} to derive an effective self-energy for the electrons in the nanowire.
%something about shell roughness <-> l_e
With this approach, the self-energy itself produces a finite subgap density of states and thus allows for leakage out of the wire into the normal part of the junction.

We model the system as sketched in Fig.~\ref{fig:modelsb}(b).
We describe the nanowire (light red region marked `NW') with a one-dimensional Bogoliubov-de Gennes Hamiltonian, $H_{\rm NW} = \frac{1}{2}\int dx\, \boldsymbol \Psi^\dagger(x) {\cal H}_{\rm NW} \boldsymbol \Psi(x)$, written in terms of the Nambu spinors $\boldsymbol \Psi(x) =[{\Psi}_{\uparrow}(x),{\Psi}_{\downarrow}(x),{\Psi}^\dagger_{\downarrow}(x),-{\Psi}^\dagger_{\uparrow}(x)]^{T}$, where the field operator ${\Psi}^\dagger_{\sigma}(x)$ creates an electron with spin $\sigma$ at position $x$.
Explicitly, we use the Hamiltonian
\begin{align}
{\cal H}_{\rm NW} = \left(-\frac{\hbar^2\partial_x^2}{2m^*} - \mu - i\alpha\partial_x\sigma_y \right)\tau_z + V_{\rm Z}\sigma_z,
\end{align}
where the Pauli matrices $\boldsymbol\sigma$ and $\boldsymbol\tau$ act in spin space and particle-hole space respectively.
Furthermore, $m^*$ is the effective mass of the electrons in the wire, $\mu$ is their chemical potential, $\alpha$ is the Rashba spin-orbit strength, and $V_{\rm Z}$ is the magnitude of the Zeeman splitting in the wire.

We discretize this Hamiltonian on $N = 100$ lattice sites and write for the retarded electronic Green function on this chain
\begin{equation}
G^R(n,m;\epsilon) = \left[\frac{1}{\epsilon-{\cal H}_{\text{NW}}-\Sigma^R_{\rm SC}(\epsilon) + i0^+}\right]_{n,m},
\label{eq:gf1}
\end{equation}
where $0^+$ is a positive infinitesimal and $\Sigma^R_{\rm SC}(\epsilon)$ is the self-energy due to the coupling to the superconductor, which we will derive below.
From this Green function we calculate the reflection matrix of the hybrid wire,
\begin{align}
R(\epsilon) &= \twomatrix{r_{\text{ee}}(\epsilon)}{r_{\text{eh}}(\epsilon)}{r_{\text{he}}(\epsilon)}{r_{\text{hh}}(\epsilon)}\nonumber\\
&=1-2i\pi W^\dag\left\{ \left[G^R(\epsilon)\right]^{-1}+i\pi WW^\dag\right\}^{-1}W,
\label{eq:refl}
\end{align}
where the amplitudes $r_{\rm ee(hh)}$ describe normal electron(hole) reflection and the off-diagonal amplitudes $r_{\rm eh,he}$ describe Andreev reflection.
The matrix
\begin{equation}
W = \sqrt{\gamma_W}\left({\bf s}_1 \otimes \mathds{1}_4\right)^{T},
\end{equation}
models the coupling between the probe lead and the first site of the chain.
Here, $\gamma_W$ parametrizes the coupling strength, $\mathds{1}_4$ is a $4\times 4$ unit matrix, and the $N$-dimensional vector ${\bf s}_1 = (1,0,0,0,\dots)$ specifies the position of the probe along the chain.
The resulting reflection matrix allows us to calculate the (zero-temperature) differential conductance as
\begin{equation}
\frac{dI}{dV} = \frac{e^2}{h}\text{Tr}\big[
1 -r_{\text{ee}}(\epsilon)^\dagger r_{\text{ee}}(\epsilon)+r_{\text{eh}}(\epsilon)^\dagger r_{\text{eh}}(\epsilon)
\big],
\label{eq:currrefl}
\end{equation}
where we set $\epsilon = eV$, in terms of the bias voltage $V$ on the tunnel probe.
More details about this calculation can be found in Ref.~\cite{Hansen2016}.

The task left is to find a suitable self-energy $\Sigma^R_{\rm SC}(\epsilon)$ that accounts for the diffusive nature of the superconductor as well as the presence of a normal drain lead.
The self-energy reads most generally
\begin{equation}
\Sigma^R_{\rm SC}(x,x';\epsilon) = \tilde t^2 G^R_{\rm SC}(x,x';\epsilon),
\end{equation}
in terms of the electronic Green functions inside the superconductor connecting the two points $x$ and $x'$ at the superconductor-wire interface (for simplicity we assumed the coupling $\tilde t$ to be local and constant).
The elements of $\Sigma^R_{\rm SC}$ thus follow straightforwardly from the electron, hole, and anomalous Green functions in the superconductor.

To find these Green functions, we assume that the SN-junction is in the dirty (diffusive) limit and that we can describe the relevant electron dynamics in the junction using a semi-classical approximation, i.e., we assume that both the electronic mean free path and the Fermi wave length in the junction are much smaller than all relevant length scales in the wire, both very reasonable assumptions.
The semi-classical Green functions then obey the Usadel equation~\cite{Eilenberger1968,Usadel1970}, which one can solve for the SN-junction, assuming no interface barrier and setting the order parameter $\Delta({\bf r})$ to a constant $-i\Delta$ inside the superconductor and to zero in the normal metal~\cite{Kuprianov1988,Belzig1996}.
By doing so, one ignores the requirement for self-consistency of $\Delta({\bf r})$, and in that sense the result must be seen as a lowest-order approximation, which is expected to introduce small quantitative errors but not to affect the result in a serious qualitative way.

The solution for the semi-classical electronic and anomalous retarded Green functions presented in Ref.~\cite{Belzig1996} uses the angular parametrization
\begin{align}
g^R_{\rm ee} (x,\epsilon) = {} & {} \cos \theta(x,\epsilon), \label{eq:gee}\\
f^R_{\rm eh} (x,\epsilon) = {} & {} \sin \theta(x,\epsilon),
\end{align}
where the position-dependent angle $\theta$ reads explicitly 
\begin{widetext}
\begin{align}
\theta(x,\epsilon) = 
\begin{cases}
4 \arctan \left\{ e^{-(x/\xi_{\rm N}) \sqrt{-i\epsilon / \Delta }} \tan \left[\frac{1}{2} \arctan \beta \right]\right\}
& \text{for } x>0, \\
4 \arctan \left\{e^{(x/\xi_{\rm S}) \sqrt[4]{1-(\epsilon/\Delta)^2}} \tan \left[\frac{1}{2} \arctan \beta +\frac{1}{4} \arctan (\Delta/i\epsilon) \right]\right\} - \arctan (\Delta/i\epsilon)
& \text{for } x \leq 0,
\end{cases}
\end{align}
assuming that the SN-interface is at $x=0$ with the superconducting region at $x \leq 0$.
We used the notation
\begin{align}
\beta = -\sin \left(\frac{1}{2} \arctan \frac{\Delta}{i\epsilon}\right) \left[\kappa \frac{ \sqrt{-i\epsilon}}{\sqrt[4]{\Delta^2-\epsilon^2}}+\cos \left(\frac{1}{2} \arctan \frac{\Delta}{i\epsilon}\right)\right]^{-1},
\end{align}
\end{widetext}
and introduced the quantities
\begin{align}
\xi_{{\rm N},{\rm S}} = {} & {} \sqrt{\hbar D_{{\rm N},{\rm S}} / 2\Delta }, \\
\kappa = {} & {} \sigma_{\rm N}\xi_{\rm S} / \sigma_{\rm S}\xi_{\rm N},
\end{align}
with $D = \frac{1}{3} v_{{\rm F}} l_{e}$ the electronic diffusion constant in terms of the electronic mean free path $l_e$, and $\sigma$ the normal-state conductivity, which both can be different in the normal and superconducting regions.

This allows us to derive straightforwardly a (position-dependent) self-energy for the electrons in the wire due to the proximity of the SN-junction,
\begin{align}
\Sigma^R_{\rm SC} (x, \epsilon) =\zeta
\big[\sin  {} & {} \theta(x,\epsilon) \tau_y 
-i \cos \theta(x,\epsilon)\big],
\label{eq:seb}
\end{align}
where the parameter $\zeta$ characterizes the coupling between the wire and the junction.
(Note that this self-energy is diagonal in the coordinate basis.)
The energy $\epsilon$ could be given a finite imaginary part, $\epsilon \to \epsilon + i \Gamma_{\rm in}$, to account for inelastic scattering processes in the diffusive junction, which would introduce an extra broadening that smears out all features in the conductance spectrum.
In all of the following, however, we disregard these processes and we set $\Gamma_{\rm in} = 0^+$.

Setting $\kappa \to 0$ corresponds to setting the conductivity of the normal part of the junction to zero.
This should effectively remove the inverse proximity effect caused by the normal part, and make the self-energy reduce to that of a bulk superconductor.
One can check that in the limit of $\kappa \to 0$ we have $\arctan (\Delta/i\epsilon) = -2\arctan\beta$, which yields $\theta (x,\epsilon) = -\arctan(\Delta/i\epsilon)$ for $x \leq 0$.
This indeed produces a position-independent self-energy $\Sigma^R_{\rm SC}(\epsilon)$ identical to the self-energy one finds for a clean bulk superconductor~\cite{PhysRevB.84.144522,Hansen2016}.
The further $\kappa$ increases, the more the self-energy deviates from this ``clean'' result.

\begin{figure}[b]
	\begin{center}
		\includegraphics[scale=1]{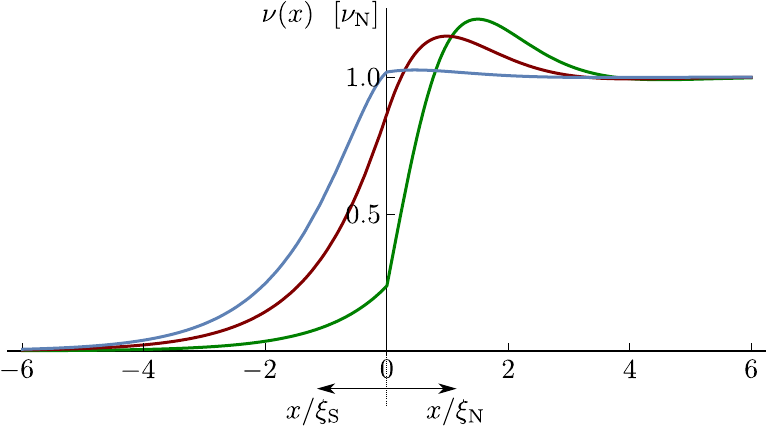}
		\caption{The position-dependent density of states $\nu(x,\epsilon)$ at energy $\epsilon = \Delta/2$ in a diffusive SN-junction, calculated from the electronic Green function (\ref{eq:gee}).  We used $\kappa = 0.2$ (green), $\kappa = 1$ (red), and $\kappa = 5$ (blue).}
		\label{fig:ndos}
	\end{center}
\end{figure}

We note that also in the limit $|x/\xi_{\rm S}| \gg 1$ the self-energy reduces to that of the clean bulk superconductor:
The coherence length $\xi_{\rm S}$ thus determines how far from the SN-interface the normal metal part still has a significant influence on the electron dynamics inside the superconductor.
To illustrate this, we calculate the position-dependent density of states in the junction, $\nu(x,\epsilon) = \nu_{\rm N}\, {\rm Re}[g^R_{\rm ee}(x,\epsilon)]$, where $\nu_{\rm N}$ is the density of states at the Fermi level in the normal state (assumed the same in the whole junction).
In Fig.~\ref{fig:ndos} we show the result at $\epsilon = \Delta/2$ for three different coupling parameters: $\kappa = 0.2$ (green), $\kappa = 1$ (red), and $\kappa = 5$ (blue).
The inverse proximity effect clearly weakens with increasing $\kappa$, but becomes always exponentially suppressed when $x$ exceeds the coherence length $\xi_{\rm S}$.

We can now calculate the differential conductance of the system using Eqs.~(\ref{eq:refl})--(\ref{eq:currrefl}), where the self-energy matrix is defined by $\big[ \Sigma^R_{\rm SC}(\epsilon) \big]_{n,m} = \Sigma^R_{\rm SC}(x_n,\epsilon) \delta_{n,m}$ at position $x_n$, corresponding to the location of site $n$ on the chain of the tight-binding discretization.

\begin{figure*}[t!]
	\begin{center}
		\includegraphics[scale=1.1]{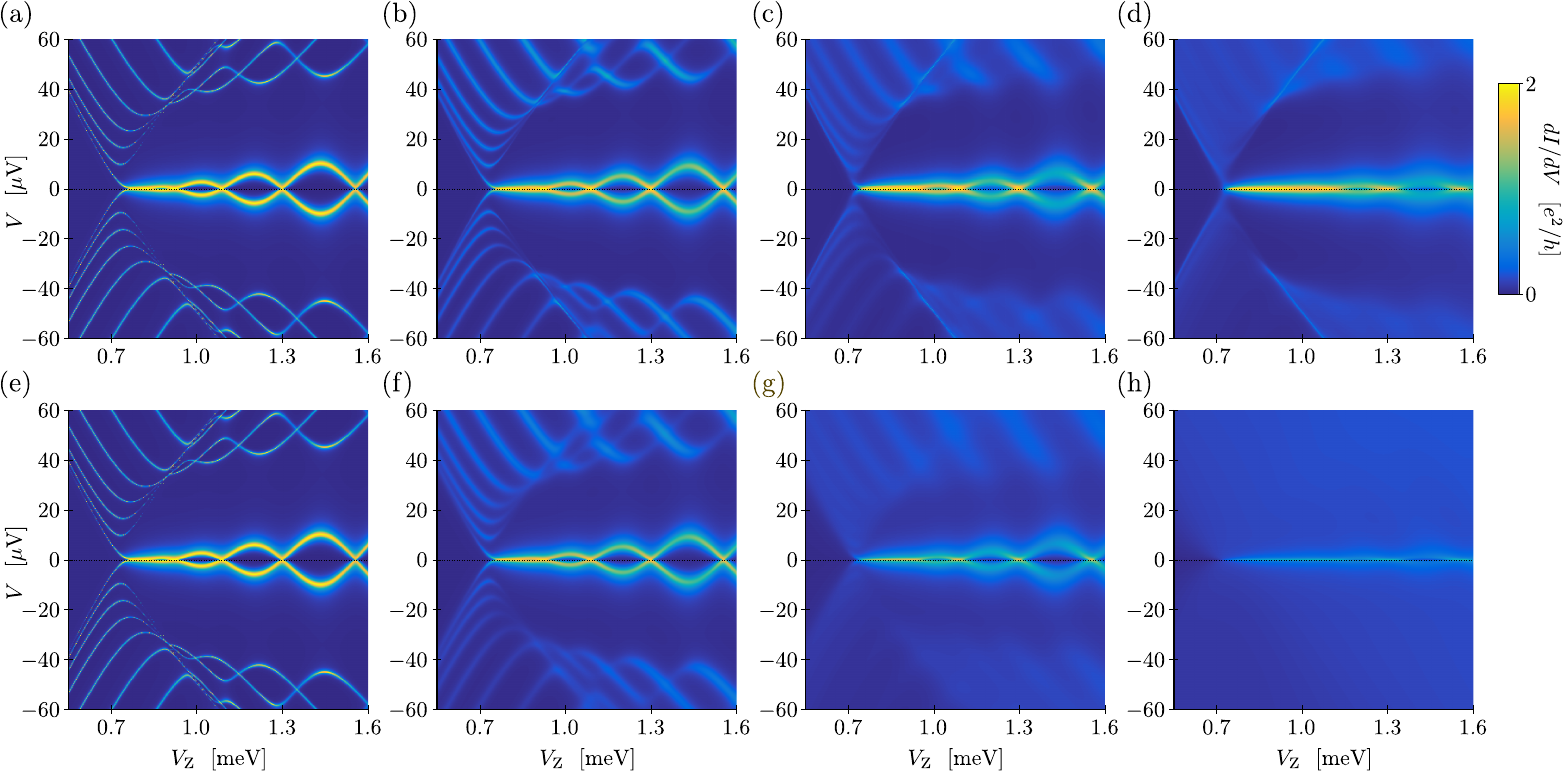}
		\caption{Calculated differential conductance as a function of Zeeman energy $V_{\rm Z}$ and applied bias voltage $V$ on the probe lead, using a self-energy that includes the effect of a (diffusive) normal metal drain coupled to the (diffusive) superconductor.
		We assumed a setup as sketched in Fig.~\ref{fig:modelsb} where we set $d=0$; all other parameters are given in the text.
		(a--d) With $L = 0.9~\mu$m and $\xi_{\rm S} = 0.1\, L$, we vary the coupling parameter $\kappa$: (a) $\kappa =0$, (b) $\kappa = 0.2$, (c) $\kappa = 1$, and (d) $\kappa = 5$.
		(e--h) We keep $L = 0.9~\mu$m and now set $\xi_{\rm S} = L$, varying again the coupling parameter $\kappa$: (a) $\kappa =0$, (b) $\kappa = 0.05$, (c) $\kappa = 0.2$, and (d) $\kappa = 1$.}
		\label{fig:plot_num3}
	\end{center}
\end{figure*}

In all numerical simulations in this section we use $m^* = 0.026\,m_e$, corresponding to the value for bulk InAs at room temperature, $\mu = 0$, $\alpha = 0.1$~eV\AA{}, $\Delta = 180~\mu$eV, $\zeta = 720~\mu$eV, and $\gamma_W = 890~\mu$eV.
The length of the wire is set to $L = 0.9~\mu$m, resulting in a lattice constant $a = 9$~nm, which is used to derive the tight-binding hopping matrix element $t = \hbar^2/2m^*a^2$ and spin-orbit-induced ``spin-flip'' nearest-neighbor coupling $s = \alpha/2a$.

For simplicity we set the distance $d$ between the SN-interface and the right end of the nanowire to zero.
In this case, we could expect different behavior depending on the parameter $\xi_{\rm S}/L$:
When $\xi_{\rm S} \lesssim L$ the drain will mainly affect the right end of the wire, and thus primarily couple to the right Majorana mode (when in the topological regime).
If, however, $\xi_{\rm S} \gtrsim L$ then we expect a stronger, more homogeneous effect which will affect both Majorana modes more equally~\footnote{This limit will also more closely resemble a situation where the superconductor is not epitaxially grown but rather deposited in bulk onto the nanowire. In this case the influence of the normal drain is not necessarily expected to be stronger on one particular side of the wire. Such a geometry was more common in the first generation of experiments.}.
For a typical experimental setup, we make the following very rough estimate:
We assume the superconductor to be epitaxial aluminum and to have a thickness of $\sim 10$~nm.
Setting $l_e = 10$~nm (assuming that the surfaces of the epitaxial layer are rough enough to randomize the electronic momentum after scattering from the surface) and using $v_{\rm F} = 2\cdot 10^6$~m/s and $\Delta = 180~\mu$eV, we find $\xi_{\rm S} \approx 100$~nm, which typically corresponds to $\xi_{\rm S}/L \sim 0.1$.

In Fig.~\ref{fig:plot_num3} we show the calculated differential conductance, as a function of applied Zeeman field $V_{\rm Z}$ and bias voltage $V$, for two different coherence lengths $\xi_{\rm S}$.
In the first four plots (a--d) we set $\xi_{\rm S} = 0.1\, L$ and we vary the coupling parameter $\kappa$: (a) $\kappa = 0$, (b) $\kappa = 0.2$, (c) $\kappa = 1$, and (d) $\kappa = 5$.
In plots (e--h) we take a longer coherence length, $\xi_{\rm S} = L$, and again vary the coupling $\kappa$: (e) $\kappa = 0$, (f) $\kappa = 0.05$, (g) $\kappa = 0.2$, and (h) $\kappa = 1$.

Roughly speaking, we see the following behavior:
(i) For $\xi_{\rm S}$ small compared to the length of the wire---where the drain is expected to couple mainly to states living at the right end of the wire---the amplitude of the Majorana oscillations is suppressed with increasing $\kappa$, and they tend to collapse to a single zero-bias peak, which then approaches $2e^2/h$ again.
(ii) For larger $\xi_{\rm S}$---where the effective coupling to the drain is more uniform across the wire---the oscillations are again suppressed, but now all features get smeared out and the conductance approaches $2e^2/h$ nowhere.
(iii) In general the gap seems to be softer at higher fields (associated with the topological regime) than in the trivial regime; this effect appears more prominent for more uniform coupling (larger $\xi_{\rm S}$).

\begin{figure*}[t]
\begin{center}
\includegraphics[scale=1.05]{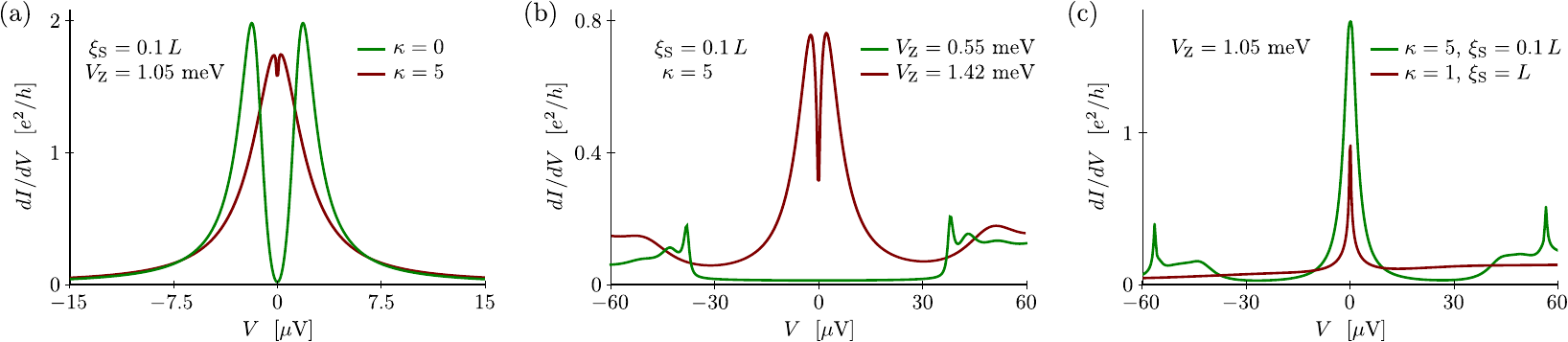}
\caption{Line cuts from the differential conductance shown in Fig.~\ref{fig:plot_num3}. All choices of parameters are indicated in the plots.}
\label{fig:cuts}
\end{center}
\end{figure*}

We illustrate this in more detail in Fig.~\ref{fig:cuts}, where we present line cuts of the data of Fig.~\ref{fig:plot_num3}.
In Fig.~\ref{fig:cuts}(a) we illustrate how the Majorana oscillations become suppressed and collapse to  a zero-bias peak:
We used $\xi_{\rm S} = 0.1\, L$ (as in the upper row in Fig.~\ref{fig:plot_num3}), fixed $V_{\rm Z} = 1.05~$meV (where the peak is clearly split at weak coupling), and show the differential conductance for $\kappa = 0$ (green) and $\kappa = 5$ (red).
At strong coupling the splitting is indeed reduced, but the peak height is still close to $2e^2/h$.
In Fig.~\ref{fig:cuts}(b), we show how the hardness of the gap can look different on opposite sides of the phase transition, when the coupling is strong:
We have set $\xi_{\rm S} = 0.1\, L$ and $\kappa = 5$, and show the conductance at $V_{\rm Z} = 0.55$~meV (green, corresponding to the trivial phase) and $V_{\rm Z} = 1.42$~meV (red, corresponding to the topological phase).
The gap is clearly less hard in the high-field case, where the zero-bias feature has a width of the same order of magnitude as the gap.
Finally, in Fig.~\ref{fig:cuts}(c) we compare the peak heights in the strong-coupling limit for $\xi_{\rm S} = 0.1\, L$ (green) and $\xi_{\rm S} = L$ (red), both at $V_{\rm Z} = 1.05$~meV.
This confirms that for longer $\xi_{\rm S}$ (a more homogeneous influence of the normal drain) not only the oscillations become suppressed, but also the actual peak heights.

\section{Analytic toy model}

We will now try to develop a better understanding of the results presented in the previous section.
To this end, we will use a simple toy model to describe the spectroscopy setup, including the influence of a normal lead connected to the proximitizing superconductor, and we will focus on the low-energy features in the spectrum (the zero-bias peak and the Majorana oscillations).

\begin{figure}[b]
	\begin{center}
		\includegraphics[scale=1.2]{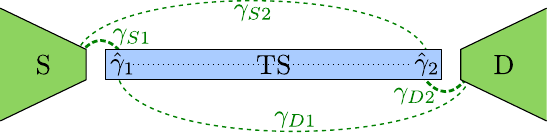}
		\caption{A toy model to include the effect of a (normal) drain lead connected to the superconductor proximitizing the nanowire. The hybrid nanowire is described in terms of a single low-energy fermionic bound state, which is split into two Majorana modes localized close to the ends of the wire.}
		\label{fig:model_a}
	\end{center}
\end{figure}

The model we will use is sketched in Fig.~\ref{fig:model_a}:
We assume that the nanowire is in the topological regime and that the gap separating the lowest-lying modes from all other states is much larger than all energy scales relevant for the dynamics of these modes.
In this case, we can project our description to this low-energy subspace and treat the hybrid wire (the blue region marked `TS') as an effective two-level system of localized Majorana modes (indicated by the two $\hat\gamma$'s).
The probe lead is tunnel coupled to the left Majorana mode, and also has a finite but weaker coupling to the right mode, due to the finite length of the wire.
The normal metal contact connected to the superconductor is modeled as a second lead which is tunnel coupled to both Majorana modes and is assumed to be at the same chemical potential as the superconductor.

To calculate the differential conductance $dI/dV$ of this system, we proceed along the same lines as in Ref.~\onlinecite{Flensberg2010}.
We first write an effective Hamiltonian
\begin{align}
\hat H = \sum_{\alpha = S,D} ( \hat H_{\alpha} + \hat H_{T,\alpha} ) + \hat H_M,
\end{align}
where
\begin{align}
\hat H_\alpha = \sum_{k,\sigma} \xi_{\alpha k\sigma} \hat c^\dagger_{\alpha k\sigma}\hat c_{\alpha k\sigma},
\end{align}
describes the electrons in lead $\alpha$,
\begin{align}
\hat H_M = \frac{i}{2} \sum_{i,j = 1,2} t_{ij} \hat \gamma_i \hat \gamma_j,
\end{align}
accounts for the coupling between the two Majorana modes, and
\begin{align}
\hat H_{T,\alpha} = \sum_{k,\sigma,i} \big( V^*_{\alpha k\sigma i} \hat c^\dagger_{\alpha k\sigma} - 
V_{\alpha k\sigma i} \hat c_{\alpha k\sigma} \big) \hat \gamma_i,
\end{align}
describes the coupling between the Majorana modes and the electrons in lead $\alpha$.
The current from lead `S' into the wire can then be expressed as
\begin{align}
I = \frac{e}{\hbar} \sum_{k,\sigma,i} \big[ V^*_{S k \sigma i} G^<_{i,S k \sigma}(0)
- V_{S k \sigma i} G^<_{S k \sigma, i}(0) \big],
\end{align}
in terms of the mixed lead-wire lesser Green functions
\begin{align}
G^<_{i,\alpha k \sigma} (t) {} & {} = i \langle \hat c^\dagger_{\alpha k \sigma}(0) \hat \gamma_i(t)\rangle, \\
G^<_{\alpha k \sigma,i} (t) {} & {} = i \langle \hat \gamma_i(0) \hat c_{\alpha k \sigma}(t)\rangle.
\end{align}
A lengthy but straightforward calculation then results in
\begin{align}
\frac{dI}{dV} = \frac{2e^2}{h} \int d\omega \, M(\omega)\,[- n'_{\rm F}(\omega - eV)],
\end{align}
where $n'_{\rm F}(x) = dn_{\rm F}(x)/dx$ is the derivative of the Fermi function, $V$ is the voltage on the probe, and the spectral density $M(\omega) = A_{SS}(\omega) + \frac{1}{2} A_{DS}(\omega) + \frac{1}{2}T_{DS}(\omega)$, with
\begin{align}
A_{\beta\alpha}(\omega) {} & {} =
{\rm Tr} \big\{ {\bf G}^{R}_\omega \boldsymbol\Gamma_\beta(-\omega)^* {\bf G}^{A}_\omega \boldsymbol\Gamma_\alpha(\omega)\big\}, \\
T_{\beta\alpha}(\omega) {} & {} =
{\rm Tr} \big\{ {\bf G}^{R}_\omega \boldsymbol\Gamma_\beta(\omega) {\bf G}^{A}_\omega \boldsymbol\Gamma_\alpha(\omega)\big\},
\end{align}
describing respectively the probabilities of Andreev reflection and normal reflection, from lead $\alpha$ to lead $\beta$ at energy $\omega$.
These probabilities are expressed in terms of the Majorana-lead coupling matrices
\begin{align}
\big[\boldsymbol\Gamma_\alpha(\omega)\big]_{ij}
= 2\pi \sum_{k \sigma} V_{\alpha k \sigma i} V^*_{\alpha k \sigma j}
\delta(\omega - \xi_{\alpha k \sigma}),
\end{align}
and the Green function of the Majorana modes,
\begin{align}
{\bf G}^{R}_\omega  = 
\frac{2}{\omega - 2i{\bf t} + i[\boldsymbol \Gamma_\omega + \boldsymbol \Gamma_{-\omega}^*]
- 2[\boldsymbol \Lambda_\omega - \boldsymbol \Lambda_{-\omega}^*]},
\end{align}
where ${\bf t}$ is the Majorana coupling matrix and we used the notation $\boldsymbol\Gamma_\omega = \boldsymbol\Gamma_S(\omega)+\boldsymbol\Gamma_D(\omega)$ and introduced the matrix $\boldsymbol\Lambda_\omega = \boldsymbol\Lambda_S(\omega)+\boldsymbol\Lambda_D(\omega)$ with
\begin{align}
\big[\boldsymbol \Lambda_\alpha(\omega)\big]_{ij} =
{\cal P} \int \frac{dz}{2\pi} \frac{\big[\boldsymbol\Gamma_\alpha(\omega)\big]_{ij}}{\omega-z}.
\end{align}

For simplicity we now assume that we can neglect, over the range of all relevant energies, all energy- and spin-dependence of both the coupling elements $V_{\alpha k \sigma i}$ and the densities of states of the leads.
This allows us to (i) simplify $[\boldsymbol \Gamma_\alpha]_{ij} = 2\pi V_{\alpha i}V^*_{\alpha j} \nu_\alpha$, where $\nu_\alpha$ is the density of states at the Fermi level of lead $\alpha$, and (ii) set $\boldsymbol\Lambda_\alpha (\omega) = 0$.
If we furthermore assume that the lead-mediated Majorana-Majorana coupling will be dominated by the overlap-induced couplings $t_{ij}$, then the coupling matrices are real and diagonal,
\begin{align}
\boldsymbol \Gamma_S = \left( \begin{array}{cc} \gamma_{S1} & 0 \\ 0 & \gamma_{S2} \end{array} \right)
\quad\text{and}\quad
\boldsymbol \Gamma_D = \left( \begin{array}{cc} \gamma_{D1} & 0 \\ 0 & \gamma_{D2} \end{array} \right),
\label{eq:gammas}
\end{align}
where $\gamma_{\alpha i}$ thus parametrizes the decay rate of Majorana mode $i$ into lead $\alpha$.
For a not too short wire one usually has $\gamma_{S2} < \gamma_{S1}$.
The magnitude of the ``leakage'' rates $\gamma_{Di}$ to the drain as well as their ratio $\gamma_{D1}/\gamma_{D2}$ depend on the actual geometry of the experimental setup, e.g.~on how far from the wire the superconductor is contacted by a normal lead, but also on the detailed electronic dynamics inside the superconductor and the normal contact.
We emphasize here that at this point these leakage rates $\gamma_{Di}$ have turned into phenomenological parameters, which do not necessarily originate from the proximity of a normal drain lead:
Leakage to any subgap density of states in the superconductor, irrespective of its origin, can be described using these model parameters.

Setting ${\bf t} = -i\sigma_y t_0 / 2$ we can now find an explicit expression for the spectral density,
\begin{align}
M(\omega) = \frac{ 4 G_{12}\omega^2 + 4G_{21}(t_0^2 + 4\Gamma_1\Gamma_2)}{\omega^4 - 2\big(t_0^2 - 2[\Gamma_1^2 + \Gamma_2^2]\big)\omega^2 +(t_0^2 + 4\Gamma_1\Gamma_2)^2},
\label{eq:m}
\end{align}
with $G_{ij} = \Gamma_1 \gamma_{Si} + \Gamma_2\gamma_{Sj}$, and using the total Majorana decay rates $\Gamma_i = \gamma_{Si} + \gamma_{Di}$.

In the limit of zero temperature (which we will assume from now on for simplicity) the differential conductance thus reads
\begin{align}
\frac{dI}{dV} = \frac{2e^2}{h} M(eV),
\label{eq:dc0}
\end{align}
with $M(\omega)$ as given in Eq.~(\ref{eq:m}).

\begin{figure}[t!]
	\begin{center}
		\includegraphics[scale=1.15]{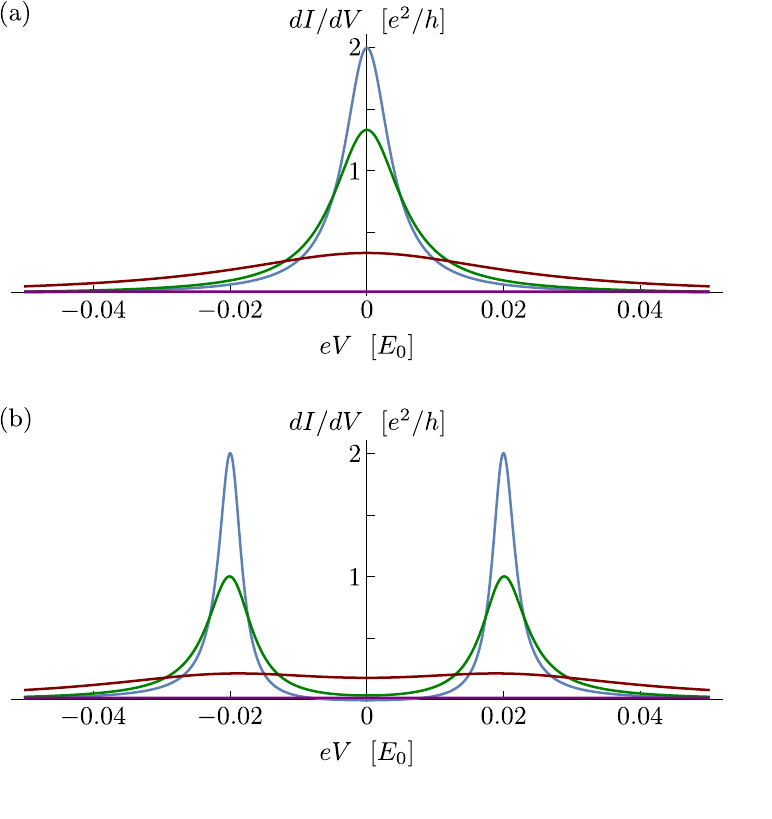}
		\caption{Differential conductance calculated using Eq.~(\ref{eq:dc0}) with $\gamma_{S1} = 0.002\, E_0$ and $\gamma_{S2} = 0$, for (a) $t_0 = 0$ and (b) $t_0 = 0.02\, E_0$.
			We have set $\gamma_{D1} = \gamma_{D2} \equiv \gamma$ and we show the resulting conductance for $\gamma = 0$ (blue), $\gamma = 0.001\, E_0$ (green), $\gamma = 0.01\, E_0$ (red), and $\gamma = 0.2\, E_0$ (purple).}
		\label{fig:anal_zbp}
	\end{center}
\end{figure}

\begin{figure*}[t!]
	\begin{center}
		\includegraphics[scale=1.1]{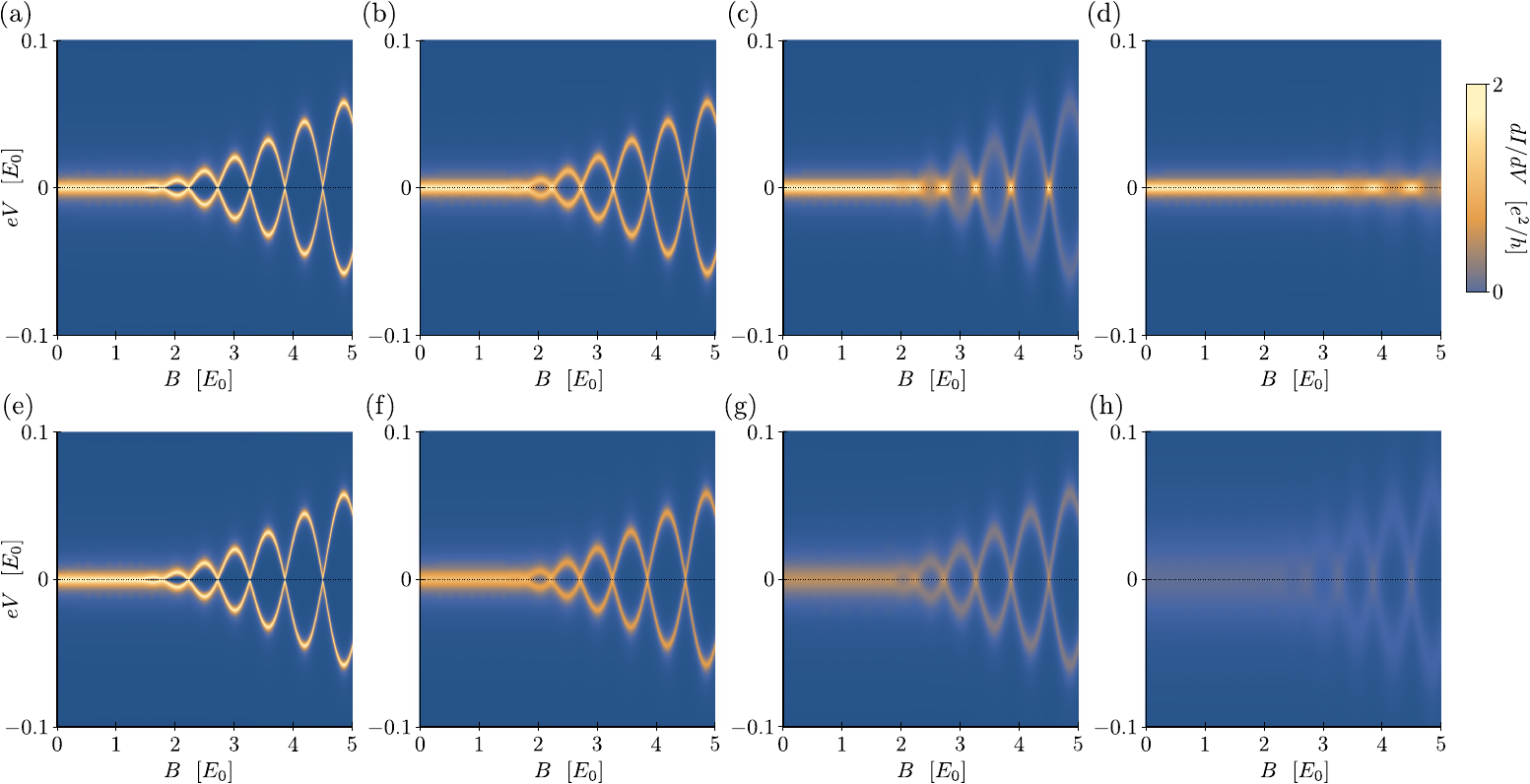}
		\caption{Differential conductance as a function of the bias voltage $V$ and the applied magnetic field $B$, calculated using Eqs.~(\ref{eq:dc0}) and (\ref{eq:m}) and using a $B$-dependent Majorana coupling energy $t_0(B)$ as given in Eq.~(\ref{eq:tb}).
		In all plots we have set $\gamma_{S1} = 0.002\, E_0$ and $\gamma_{S2} = 0$.
		(a--d) We set $\gamma_{D1} = 0$ and vary the coupling of the Majorana mode 2 to drain: (a) $\gamma_{D2} = 0$, (b) $\gamma_{D2} = 0.001\, E_0$, (c) $\gamma_{D2} = 0.01\, E_0$, and (d) $\gamma_{D2} = 0.2\, E_0$.
		(e--h) We set $\gamma_{D1} = \gamma_{D2} \equiv \gamma$ and vary this uniform coupling parameter: (a) $\gamma = 0$, (b) $\gamma = 0.001\, E_0$, (c) $\gamma = 0.003\, E_0$, and (d) $\gamma = 0.01\, E_0$.}
		\label{fig:anal_plot}
	\end{center}
\end{figure*}

Let us first investigate this result in the ideal case of a wire that is long enough so that $t_0$ and $\gamma_{S2}$ can be neglected.
In that limit, the differential conductance,
\begin{align}
\frac{dI}{dV} = \frac{2e^2}{h} \frac{4\gamma_{S1}\Gamma_1}{4\Gamma_1^2 + (eV)^2},
\label{eq:lorentz}
\end{align}
acquires a Lorentzian line shape with a full width half maximum of $4\Gamma_1$ and a maximum conductance of $(2e^2/h)(\gamma_{S1}/\Gamma_1)$ at $V = 0$.
We see that this zero-bias peak has a height of $2e^2/h$ as long as $\gamma_{D1}=0$, i.e., as long as Majorana mode 1 (which is completely decoupled from mode 2) does not have a second normal channel into which it can decay.
When $\gamma_{D1}$ becomes finite, e.g.~due to a superconductor-mediated coupling to a normal drain lead, the peak height is suppressed by a factor $\gamma_{S1}/\Gamma_1$.
We illustrate this in Fig.~\ref{fig:anal_zbp}(a), where we plot the differential conductance as given by (\ref{eq:dc0}) with $t_0 = 0$, $\gamma_{S2} = 0$, and $\gamma_{S1} = 0.002\, E_0$, in terms of the (arbitrary) energy scale $E_0$.
We varied the coupling of Majorana mode 1 to the drain lead as $\gamma_{D1} = 0$ (blue), $\gamma_{D1} = 0.001\, E_0$ (green), $\gamma_{D2} = 0.01\, E_0$ (red), and $\gamma_{D1} = 0.2\, E_0$ (purple).

Another idealized limit is where the wire is short enough that $t_0$ cannot be neglected, but the probe lead is still coupled to mode 1 only and there is no effective leakage to a second normal contact at all, i.e.~$\gamma_{S2} = \gamma_{D1} = \gamma_{D2} = 0$.
In that case, Eq.~(\ref{eq:dc0}) gives
\begin{align}
\frac{dI}{dV} = \frac{2e^2}{h} \frac{4\gamma_{S1}^2(eV)^2}{t_0^4 - 2(t_0^2 - 2\gamma_{S1}^2)(eV)^2 + (eV)^4},
\end{align}
which yields zero conductance at $V=0$ and produces two peaks at $V = \pm t_0/e$ with a height of $2e^2/h$ and full width half maximum of $2\gamma_{S1}$, as expected.
If we now add a finite coupling to the drain, the double-peak structure gets suppressed.
This is illustrated in Fig.~\ref{fig:anal_zbp}(b), where we again set  $\gamma_{S1} = 0.002\, E_0$ and $\gamma_{S2} = 0$, but now use $t_0 = 0.02\, E_0$.
The leakage rates to the drain are set equal, $\gamma_{D1} = \gamma_{D2} \equiv \gamma$, and are gradually increased from $\gamma = 0$ (blue), $\gamma = 0.001\, E_0$ (green), $\gamma = 0.01\, E_0$ (red), to $\gamma = 0.2\, E_0$ (purple).

In general, the function $M(eV)$ can have a single- or double-peak structure, depending on the relative magnitude of the Majorana coupling energy $t_0$ and the decay rates $\gamma_{\alpha i}$.
For a given set of parameters, we find a single peak at $V=0$ when
\begin{align}
%t_0^2 < t^2_c \equiv 4\frac{\Gamma_1^3\gamma_{S2} + \Gamma_2^3\gamma_{S1} }{2 G_{21}+G_{12}}.
t_0^2 < t^2_c \equiv 4\frac{\Gamma_1^3\gamma_{S2} + \Gamma_2^3\gamma_{S1} }{\Gamma_1(\gamma_{S1} + 2 \gamma_{S2}) + \Gamma_2(2\gamma_{S1} + \gamma_{S2}) },
\label{eq:tc}
\end{align}
and the peak splits into two when $t^2_0>t^2_c$.
We see that $t_c$ being non-zero depends on having a finite decay rate of Majorana mode 2: as long as $\Gamma_2 = 0$, which implies that $\gamma_{S2} = \gamma_{D2} = 0$, we have $t_c=0$.
From Eq.~(\ref{eq:tc}) it is thus clear how the effective coupling of the Majorana modes to normal metal leads---either to the tunnel probe itself or to a normal drain connected to the superconductor---can suppress the split-peak behavior of the conductance at finite $t_0$.
In the case of a proximitized semiconductor nanowire, where the effective coupling between the two low-energy modes is expected to oscillate as a function of the applied magnetic field, leakage to a drain lead can thus obscure the corresponding oscillatory pattern in the differential conductance.

To connect to the numerical results presented in Fig.~\ref{fig:plot_num3}, we show in Fig.~\ref{fig:anal_plot} the conductance spectrum, as calculated from Eq.~(\ref{eq:dc0}), as a function of both the bias voltage $V$ and applied Zeeman field $B$.
The effect of the field is implemented phenomenologically by setting
\begin{align}
	t_0(B) = \frac{E_0}{\sqrt b} e^{-l /2b} \cos\big(l \sqrt b \big),
	\label{eq:tb}
\end{align}
which corresponds to the coupling between the two Majorana modes due to the finite overlap of their wave functions, assuming for simplicity the large-$B$ limit~\cite{DasSarma2012,Rainis2013}.
In this expression the Zeeman field $b = B/E_0$ is expressed in units of the (arbitrary) energy scale $E_0$ and the parameter $l$ characterizes the length of the wire.
By setting $l = L\sqrt{2m E_0}/\hbar$ (with $L$ the actual length of the wire) and $E_0 = (2m\alpha^2\Delta^2/\hbar^2)^{1/3}$ (with $\alpha/\hbar$ the spin-orbit velocity and $\Delta$ the induced superconducting gap) one can connect these parameters to those used in more detailed models describing a semiconducting spin-orbit coupled nanowire in proximity to a superconductor~\cite{DasSarma2012}.
We further assume that the wire is long enough so that we can neglect $\gamma_{S2}$ and we investigate again the two different situations corresponding to the two rows in Fig.~\ref{fig:plot_num3} (corresponding to different coherence lengths $\xi_{\rm S}$).

In Fig.~\ref{fig:anal_plot}(a--d) we set $\gamma_{D1} = 0$ and vary the coupling $\gamma_{D2}$ of the right Majorana mode to the drain lead.
This situation is expected to be more close to that of the top row of Fig.~\ref{fig:plot_num3}, where the coherence length $\xi_{\rm S} = 0.1\, L$ was significantly smaller than the wire length, and the normal drain thus mainly affected states with most of their weight close to the right end of the wire.

We used (a) $\gamma_{D2} = 0$, (b) $\gamma_{D2} = 0.001\, E_0$, (c) $\gamma_{D2} = 0.01\, E_0$, and (d) $\gamma_{D2} = 0.2\, E_0$, and
we see that increasing $\gamma_{D2}$ again affects the appearance of the Majorana oscillations:
For intermediate $\gamma_{D2}$ the split-peak structure gets smeared out, and for large $\gamma_{D2}$
ultimately all oscillatory behavior gets suppressed and the conductance spectrum shows a single zero-bias peak that approaches $2e^2/h$, in a way very similar to what we found numerically in Fig.~\ref{fig:plot_num3}(a--d).
To check, we evaluate from (\ref{eq:tc}) the corresponding values for the critical Majorana coupling $t_c$, and find (a) $t_c = 0$, (b) $t_c = 0.001\, E_0$, (c) $t_c \approx 0.013\, E_0$, and (d) $t_c \approx 0.28\, E_0$.
These values are indeed consistent with the global behavior seen in Fig.~\ref{fig:anal_plot}:
The regions in Fig.~\ref{fig:anal_plot}(b--d,f--h) where $t_0(B) < t_c$, i.e., where the original splitting observed in Fig.~\ref{fig:anal_plot}(a,e) is smaller than $t_c$, indeed seem to be the regions where we see a single zero-bias peak instead of a split peak.

We can understand the behavior observed in Fig.~\ref{fig:anal_plot}(a--d) as follows:
If Majorana mode 2 is the only mode coupled to the drain lead and the corresponding decay rate presents the largest energy scale in the model, then this rapid decay will prevent the two Majorana modes from hybridizing:
All coherence built up between the two modes (which happens on the time scale $\hbar/t_0$) decays before it becomes significant (the decay happens on the time scale $1/\gamma_{D2}$).
The Majorana modes thus become decoupled from each other if the decay is fast enough, and this brings us effectively to the situation where $t_0$ is zero and each Majorana mode is only coupled to the nearest lead.
This explains why a single zero-bias peak of height $2e^2/h$ is recovered in Fig.~\ref{fig:anal_plot}(a--d) in the regime where $\gamma_{D2} \gtrsim t_0$.

In Fig.~\ref{fig:anal_plot}(e--h) we try to reproduce the situation of the lower row of Fig.~\ref{fig:plot_num3}, where a longer coherence length $\xi_{\rm S} = L$ resulted in a more homogeneous influence of the drain lead.
To mimic this, we make the two couplings to the drain equal, $\gamma_{D1} = \gamma_{D2} \equiv \gamma$, and vary the parameter $\gamma$, using (e) $\gamma = 0$, (f) $\gamma = 0.001\, E_0$, (g) $\gamma = 0.003\, E_0$, and (h) $\gamma = 0.01\, E_0$.
The results now indeed resemble more those shown in the lower row of Fig.~\ref{fig:plot_num3}:
Increasing $\gamma$ not only quenches the Majorana oscillations, but also suppresses the height of all conductance features.
This results from Majorana mode 1 (the one probed by the source) now also being coupled to the drain and thus acquiring an effective life-time broadening.

In conclusion, our simple low-energy toy model thus shows how a finite leakage rate from the Majorana modes into a normal metal drain can suppress the Majorana oscillations in the conductance spectrum.
The result is either a persistent zero-bias peak or an overall broadening and suppression of all features in the spectrum, depending on the ratio of the leakage rates of the two modes.
Similar phenomena are indeed commonly observed in experiments, even in very recent high-quality spectroscopy experiments~\cite{Deng2016,Zhang2016}.

\section {Conclusion}

Tunneling spectroscopy experiments on proximitized semiconducting nanowires---aimed at the detection of convincing signatures of zero-energy Majorana modes---inevitably involve a normal drain lead connected to the hybrid system.
In this paper, we reported a detailed theoretical investigation of the effects on the conductance spectrum of the system in the presence of such a drain.
We first presented numerical results, where we treated the superconductor and the drain lead together as one diffusive SN-junction, which showed how several commonly observed ``inconsistencies'' between experimental results and theoretical predictions could, at least partly, be explained in terms of finite leakage out of the low-energy states inside the hybrid wire due to the presence the normal drain.
We then supported these results with the investigation of a simple toy model, which allowed us to derive analytic expressions for the conductance through the low-energy modes in the topological regime.
These analytic results reproduced qualitatively our main numerical findings and thus provided more insight in the physics underlying the observed phenomena.

\section{Acknowledgments}

The work was supported by the Deutsche Forschungsgemeinschaft (Bonn) within the network CRC TR 183 and by the Danish National Research Foundation. KF acknowledges the hospitality of the Heinrich-Heine University D\"{u}sseldorf within the DFG Mercator program.

%\bibliographystyle{../../../paperbib}
%\bibliography{../../../library}

\end{document}